\documentclass[%
prd,
reprint,
nodate,
superscriptaddress,
nofootinbib,
 amsmath,
 amssymb,
 aps,
]{revtex4-2}

\pdfoutput=1

\usepackage[utf8]{inputenc} 

\usepackage{xcolor}
 
\usepackage{graphicx}
\usepackage{dcolumn}
\usepackage{bm}
\usepackage{microtype}
\usepackage{hyperref}

\newcommand{\be}{\begin{eqnarray}}
\newcommand{\ee}{\end{eqnarray}}

\begin{document}

\author{Michal P.\ Heller} \email{michal.p.heller@aei.mpg.de}
\affiliation{Max Planck Institute for Gravitational Physics (Albert Einstein Institute),
14476 Potsdam-Golm, Germany}
\affiliation{National Centre for Nuclear Research, 02-093 Warsaw, Poland}

\author{Alexandre Serantes}
\email{alexandre.serantesrubianes@ncbj.gov.pl}
\affiliation{National Centre for Nuclear Research, 02-093 Warsaw, Poland}

\author{Micha\l\ Spali\'nski}
\email{michal.spalinski@ncbj.gov.pl}
\affiliation{National Centre for Nuclear Research, 02-093 Warsaw, Poland}
\affiliation{Physics Department, University of Bia{\l}ystok,
  15-245 Bia\l ystok, Poland}

\author{\\Viktor Svensson}
\email{viktor.svensson@aei.mpg.de}
\affiliation{National Centre for Nuclear Research, 02-093 Warsaw, Poland}
\affiliation{Max Planck Institute for Gravitational Physics (Albert Einstein Institute),
14476 Potsdam-Golm, Germany}

\author{Benjamin Withers}
\email{b.s.withers@soton.ac.uk}
\affiliation{Mathematical Sciences and STAG Research Centre, University of Southampton, Highfield, Southampton SO17 1BJ, UK}

\title{Hydrodynamic gradient expansion in linear response theory}

\begin{abstract}
A foundational question in relativistic fluid mechanics concerns the properties of the hydrodynamic gradient expansion at large orders. We establish the precise conditions under which this gradient expansion diverges for a broad class of microscopic theories admitting a relativistic hydrodynamic limit, in the linear regime.
Our result does not rely on highly symmetric fluid flows utilized by previous studies of heavy-ion collisions and cosmology.
The hydrodynamic gradient expansion diverges whenever energy density or velocity fields have support in momentum space exceeding a critical momentum, and converges otherwise.
This critical momentum is an intrinsic property of the microscopic theory and is set by branch point singularities of hydrodynamic dispersion relations.
\end{abstract}

\maketitle

\section{Introduction}

The goal of relativistic hydrodynamics is to provide an effective description of long-lived, long wavelength excitations -- hydrodynamic modes -- which are generally expected to dominate nonequilibrium dynamics of collective states of quantum fields at macroscopic scales and sufficiently late times~\cite{Kovtun:2012rj,Hartnoll:2016apf,Florkowski:2017olj,Romatschke:2017ejr}. Understanding what exact scales and times these are has been a very active field of research of the past decade in connection with studies of collective phases of strong interactions in nuclear collisions at the Relativistic Heavy Ion Collider and the Large Hadron Collider, where relativistic hydrodynamics is the framework translating between the observed particle spectra and microscopic features such as characteristics of the initial state~\cite{Heinz:2013th,Busza:2018rrf}. Related recent developments in relativistic hydrodynamics go well beyond the realm of nuclear physics and extend also to astrophysics~\cite{Shibata:2017jyf, Alford:2017rxf, Bemfica:2019cop}, as well as to studies of strong gravity~\cite{Caldarelli:2008mv, Green:2013zba}.

Much progress occurred recently thanks to the effective field theory perspective formulated as a spacetime derivative expansion~\cite{Baier:2007ix} as well as by using insights from linear response theory~\cite{Kovtun:2005ev}. The effective field theory approach expresses expectation values of conserved currents in terms of derivatives of local classical fields such as the energy density and fluid velocity. 
The energy momentum tensor is
represented as a sum of all possible terms graded by the number of derivatives, starting with the perfect fluid contribution. By comparing this formal series to the analogous gradient expansion calculated in a microscopic theory one can express the parameters appearing in the hydrodynamic series -- transport coefficients -- in terms of microscopic quantities. Interestingly, the gradient series evaluated on a solution of the evolution equations can have a vanishing radius of convergence at least in the case of highly-symmetric flows describing rapidly expanding matter, as was discovered in holography~\cite{Heller:2013fn, Buchel:2016cbj,Baggioli:2018bfa,Buchel:2018ttd,Aniceto:2018uik}, hydrodynamic models~\cite{Heller:2015dha,Basar:2015ava,Aniceto:2015mto} and kinetic theory~\cite{Denicol:2016bjh,Heller:2016rtz, Heller:2018qvh,Denicol:2019lio}.

At the linearized level~\cite{1963AnPhy..24..419K},
the dynamics is governed 
by sums of harmonic contributions with complex frequencies which encode Fourier space singularities of retarded correlators. Imaginary parts of these frequencies capture effects of dissipation. Terms associated with frequencies 
which vanish at small momentum correspond to shear and sound hydrodynamic modes, while the rest represents transient phenomena~\cite{Florkowski:2017olj}. 
Their dispersion relations are usually expressed as expansions in spatial momenta. Correspondingly, we choose the position space gradient expansion to be expressed in purely spatial derivatives, which can always be done. With this choice, the position space gradient expansion of the hydrodynamic constitutive relations is related to the shear and sound mode frequencies expanded in small spatial momentum. In \cite{Withers:2018srf} and later in \cite{Grozdanov:2019kge, Grozdanov:2019uhi} it was observed that the latter series have a finite non-zero radius of convergence, which reflects the presence of nonhydrodynamic modes. This parallels the fact that the Borel transform of the gradient expansion in an expanding plasma similarly reveals information about the nonhydrodynamic sectors.

The present article combines these two lines of research, which allows one to make for the first time generic statements about the convergence of the hydrodynamic gradient expansion across microscopic theories and models. In particular, we show that the convergence of the position space gradient expansion of the constitutive relations in the linearized regime is governed by the same mechanism that yields a finite radius of convergence of series expansions of hydrodynamic mode frequencies at small momentum: 
the radius of convergence of the dispersion relations is precisely the momentum scale which defines which flows possess a convergent asymptotic gradient expansion and which ones do not.

\section{Hydrodynamics} The expectation value of the energy-momentum tensor can be expressed as the perfect-fluid part plus corrections
\begin{equation}
\label{eq.Tmunudef}
\langle T^{\mu \nu} \rangle = \left({\cal E} + {\cal P} \right) U^{\mu} U^{\nu} + {\cal P} \, g^{\mu \nu} + \Pi^{\mu \nu}.
\end{equation}
In hydrodynamics, the correction $\Pi^{\mu \nu}$ is represented in terms of derivatives of the hydrodynamic fields which we take as the energy density $\cal E$ and flow velocity~$U^{\mu}$ with $U\cdot U = -1$. The pressure $\cal P$ is related to $\cal E$ via an equation of state~\cite{Florkowski:2017olj,Romatschke:2017ejr}.

We consider flat $d$-dimensional spacetime and use the Landau frame where $U_{\mu} \Pi^{\mu \nu} = 0$. We focus on conformal and parity-invariant theories. Conformal symmetry forces $\Pi^{\mu}_{\,\,\mu} = 0$ and ${\cal P} = {\cal E}/\left(d-1\right)$. 
The most general hydrodynamic $\Pi^{\mu \nu}$ takes now the form \cite{Grozdanov:2015kqa,Diles:2019uft}
\begin{align}
\label{eq.Pihydrodef}
\Pi^{\mu \nu} &= - \eta \, \sigma^{\mu \nu} + \tau_{\pi } \eta \, {\cal D} \sigma^{\mu \nu} - \nonumber \\
& - \frac{1}{2} \theta_1 \, {\cal D}_{\alpha} {\cal D}^{\alpha} \sigma^{\mu \nu} - \theta_2 \, {\cal D}^{\langle \mu} {\cal D}^{\nu \rangle} {\cal D}_\alpha U^\alpha  +  \ldots \, ,
\end{align}
where the ellipsis denotes terms higher than third order and we display only terms surviving linearization. The angle-brackets in \eqref{eq.Pihydrodef} denote the tensors made symmetric, transverse and traceless, ${\cal D} = U^\mu \partial_\mu$ and ${\cal D}^\mu= \left(g^{\mu\nu} + U^{\mu} U^{\nu}\right) \partial_\nu$ are respectively a comoving and a transverse derivative, $\sigma^{\mu \nu} = 2{\cal D}^{\langle \mu} U^{\nu \rangle}$ denotes the shear tensor and $\eta$ is the shear viscosity, $\tau_{\pi}$~the relaxation time and $\theta_1$, $\theta_2$ are third order transport coefficients.

We focus on small perturbations away from thermal equilibrium, i.e., we consider 
\begin{equation}
U^\mu = (1,\textbf{u})^{\mu} \quad \mathrm{and} \quad \mathcal{E}= \mathcal{E}_0 + \epsilon
\end{equation}
with $|\epsilon/{\cal E}_{0}|,|u_l u^{l}| \ll 1$. We denote spatial indices with Latin letters and spatial vectors with bold font. It is useful to work in Fourier space 
with a plane-wave Ansatz 
\begin{subequations}
\label{eq.uepsFdef}
\begin{eqnarray}
u^i(t,\textbf{x}) &=&\hat u^i(\textbf{k}) \, e^{-i \, \omega \, t + i \, \textbf{k}\cdot\textbf{x}},\\
\epsilon(t,\textbf{x}) &=& \hat\epsilon(\textbf{k}) \, e^{-i \, \omega \, t + i \, \textbf{k}\cdot\textbf{x}}. 
\end{eqnarray}
\end{subequations}
The perturbations can be decomposed into shear and sound channel components~\cite{Kovtun:2012rj}, labelled here by $\perp$ and~$\parallel$ subscripts. They are given by
\be
\hat{\textbf{u}}_\parallel = \frac{\textbf{k} \cdot \hat{\textbf{u}}}{\textbf{k}^2} \, \textbf{k}, \quad \quad
\hat{\textbf{u}}_\perp= \hat{\textbf{u}} - \hat{\textbf{u}}_\parallel. 
\ee
\normalsize
with $\hat\epsilon = 0$ vanishing in the shear channel. With no loss of generality, due to rotational invariance, we take 
\begin{equation}
\label{eq.xkconvensions}
\textbf{k} = (0,\ldots,0,k).
\end{equation}
Conservation of the energy-momentum tensor together with the hydrodynamic constitutive relation~\eqref{eq.Pihydrodef} 
determines the frequencies $\omega$ appearing in \eqref{eq.uepsFdef} as functions of~$k$. The dispersion relations 
take the form~\cite{Grozdanov:2015kqa,Diles:2019uft}
\begin{align}
&\tilde\omega_{\perp} = - i\frac{\eta}{s \, T} k^2 - i \left(\frac{\eta^2 \, \tau_\pi}{s^2 T^2} - \frac{\theta_1}{2 \, s \, T}\right) k^4 + \ldots, \nonumber \\ 
&\tilde\omega_{\parallel}^\pm = \pm c_s \, k - i \Gamma k^2 \mp \frac{\Gamma}{2\, c_s}\left(\Gamma - 2\,c_s^2\,\tau_\pi\right) k^3 - \nonumber \\
& i \left(2\,\Gamma^2\,\tau_\pi - \frac{(d-2)\,(\theta_1 + \theta_2)}{2\,(d-1)\,s\,T} \right) k^4 + \ldots \, , \label{eq.omegahs}
\end{align}
where the tilde means that these are frequencies in the hydrodynamic theory rather than in a microscopic theory. Here $T$ and $s$ are the temperature and entropy density associated with~${\cal E}_{0}$, $c_s = 1/\sqrt{(d-1)}$ is the speed of sound, and $\Gamma = (d-2)/(d-1) \eta/(s T)$.

Calculations in holography~\cite{Withers:2018srf, Grozdanov:2019kge, Grozdanov:2019uhi} reveal that the series~\eqref{eq.omegahs} have a finite and non-zero radius of convergence, with evidence going back to the studies of causal second order hydrodynamics in \cite{Baier:2007ix}. In physically interesting cases, linear response theory shows that apart from the hydrodynamic modes, there are additional excitations that are short-lived, i.e. whose 
complex frequency $\omega(k)$ has a non-vanishing imaginary part even as $k\rightarrow0$ \cite{Kovtun:2005ev,Baier:2007ix,Romatschke:2015gic,Grozdanov:2016vgg}. Explicit calculations in several representative cases show  that the radius of convergence of hydrodynamic dispersion relations is set by the magnitude $k_{*}$ of a (possibly complex) momentum for which the frequency of a hydrodynamic mode coincides with that of a nonhydrodynamic one at a branch point of $\omega(k)$ \cite{Withers:2018srf, Grozdanov:2019kge, Grozdanov:2019uhi}.

\section{Constitutive relations} To exploit the known properties of $\omega(k)$, we parametrize the gradient expansion of $\Pi^{\mu \nu}$ using only spatial derivatives. A different choice is explored in Appendix \ref{appendix_b}. 

The most general form of $\Pi^{\mu \nu}$ can be constructed from three elementary tensorial structures that are respectively 
\begin{subequations}
\begin{align}
    \sigma_{jl} &= \left(\partial_j u_l + \partial_l u_j - \frac{2}{d-1} \delta_{jl} \partial_r u^r \right), \\
    \pi_{jl}^\epsilon &= \left(\partial_j \partial_l - \frac{1}{d-1} \delta_{jl} \partial^2 \right) \epsilon, \\
    \pi_{jl}^u &= \left(\partial_j \partial_l - \frac{1}{d-1} \delta_{jl} \partial^2 \right) \partial_r u^r. 
\end{align}
\end{subequations}
The last of these appears already in \cite{Bu:2014sia,Bu:2014ena} (see also \cite{Grozdanov:2019uhi}).\footnote{\cite{Bu:2014sia,Bu:2014ena} are part of a program studying the stress tensor of holographic theories in the linear response regime (see also \cite{Lublinsky:2009kv, Bu:2015ika,Bu:2015bwa}). In contrast, we study hydrodynamic constitutive relations perturbatively in spatial gradients, and without restriction to a particular microscopic model.} With no loss of generality we write the constitutive relations as
\begin{equation}
\Pi_{jl} = -A(\partial^2) \, \sigma_{jl} - B(\partial^2) \, \pi_{jl}^u - C(\partial^2) \, \pi_{jl}^\epsilon  \label{const_rel_grad_exp}
\end{equation}
and $\Pi_{tt}=\Pi_{ti}=0$. $A, B$ and $C$ are infinite series in spatial Laplacians,
\begin{align}
\label{eq.defan}
A = \sum_{n=0}^\infty  a_n \left(-\partial^2\right)^n,
\end{align}
and the $a_n$ are transport coefficients, with similar expressions for $B$ and $C$ involving transport coefficients $b_{n}$ and~$c_{n}$. In principle, $A$, $B$ and $C$ could also depend on $\partial_t$, but in the hydrodynamic gradient expansion one can use the conservation equations to replace temporal derivatives by spatial ones in a systematic way.\footnote{See Appendix A of \cite{Grozdanov:2015kqa}.} 

It follows from \eqref{const_rel_grad_exp} that each even order in gradients introduces one new transport coefficient, while each odd order higher than one introduces two. We find it remarkable that such a simple argument implies that the number of independent transport coefficients at a given order in the gradient expansion of linearized hydrodynamics does not grow with the order, but is limited.

An analogous situation occurs in the series expansions of $\omega_\perp$, $\omega_\parallel^\pm$ around $k=0$. Since $\omega_\parallel^+$, $\omega_\parallel^-$ obey the relation $\omega_\parallel^+(k) = - \omega_\parallel^-(k)^*$, their series coefficients are not independent. These coefficients are real for odd powers of~$k$, and purely imaginary for even powers of~$k$. $\omega_\perp$ is given by a series expansion in $k^2$ with purely imaginary coefficients. Therefore, each even order in \eqref{eq.omegahs} introduces two new real parameters, while each odd order introduces just one. This counting matches the number of independent transport coefficients in \eqref{const_rel_grad_exp}, and suggests that it is possible to express $a_n$, $b_n$ and $c_n$, see \eqref{eq.defan}, in terms of the hydrodynamic dispersion relations~\eqref{eq.omegahs}.

\section{Matching} We now show explicitly that there is a direct relation between $A$, $B$ and $C$ defined in \eqref{const_rel_grad_exp} and the hydrodynamic dispersion relations~\eqref{eq.omegahs}.

For the shear mode, with the wave vector choice we made in~\eqref{eq.xkconvensions}, the only non-zero components of $\sigma_{jl}$ are 
\begin{equation}
\sigma_{1,d-1} = \sigma_{d-1,1} = i \, k \, u_1
\end{equation}
where we have taken $\textbf{u} = (u_1,0,\ldots,0)$ with no loss of generality due to rotational invariance. $\pi_{jl}^u$ and $\pi_{jl}^\epsilon$ vanish identically for this mode since $\partial_i u^i = \epsilon = 0$. 

The conservation of the energy-momentum tensor~\eqref{eq.Tmunudef}, in combination with the hydrodynamic constitutive relation~\eqref{const_rel_grad_exp}, predicts the following dispersion relation 
\begin{equation}
\tilde\omega_\perp(k) = -i\frac{1}{s \, T} \sum_{n=0}^\infty a_{n} k^{2n+2}.  
\end{equation}
Demanding that $\tilde\omega_\perp(k)$ agrees with the microscopic shear hydrodynamic mode $\omega_\perp$ at every order in an expansion around $k^2=0$ fixes the $a_n$ coefficients to be 
\begin{equation}
\label{eq.an}
a_{n}  = [k^{2n+2}]\left( i \, s\, T \, \omega_\perp\right),
\end{equation}
where the notation $[k^{p}]\left(f\right)$ denotes the coefficient of $k^{p}$ in the series expansion of $f$ around~$k = 0$.

With $A(\partial^2)$ fixed, we determine $B(\partial^2)$ and $C(\partial^2)$ by considering the sound mode. Now $\textbf{u} = (0,\ldots,0,u_{d-1})$, $\epsilon \neq 0$ and 
\begin{equation}
\pi_{jl}^u = - \frac{1}{2} k^2 \sigma_{jl}. 
\end{equation}
Furthermore, the only non-zero components of $\sigma_{jl}$ and $\pi^\epsilon_{jl}$ are
\begin{subequations}
\begin{align}
&\sigma_{jj} = - \frac{2}{d-1} \, i \, k \, u_{d-1},\,\,\,j=1...d-2,\\
&\sigma_{d-1,d-1} = \frac{2\,(d-2)}{d-1} \, i\, k\, u_{d-1} \\
&\pi^\epsilon_{jj} = \frac{1}{d-1} \, k^2 \, \epsilon,\,\,\,j=1...d-2,\\
&\pi^\epsilon_{d-1,d-1} = - \frac{d-2}{d-1} \, k^2\, \epsilon. 
\end{align}
\end{subequations}
In the end, the conservation equations reduce to 
\begin{subequations}
\label{eq.sc_cons}
\begin{align}
&- i \, \omega \, \epsilon +  i \, k\, s\, T\, u_{d-1} = 0, \label{sc_cons_1}\\
&- i\, \omega\, s\, T \, u_{d-1} + \frac{1}{d-1} \, i \, k \,\epsilon + \nonumber  \\ 
&+ \frac{d-2}{d-1}\,\sum_{n=0}^\infty (2\,a_n - b_{n-1})\, k^{2n+2} \, u_{d-1} + \nonumber  \\ 
&+ \frac{d-2}{d-1} \,\sum_{n=0}^\infty i \,c_n\, k^{2n+3}\,\epsilon = 0, \label{sc_cons_2}
\end{align}
\end{subequations}
where we have introduced $b_{-1}\equiv 0$ for brevity.
Note that the conservation equation \eqref{sc_cons_1} does not depend on transport coefficients as a result of our frame choice. \eqref{eq.sc_cons} has two
solutions, $\tilde\omega_\parallel^+(k)$ and $\tilde\omega_\parallel^-(k)$, given as series expansions around $k=0$, whose coefficients depend on $a_n, b_n$ and $c_n$. Demanding that these quantities agree with the microscopic sound modes $\omega_\parallel^+(k)$ and $\omega_\parallel^-(k)$, the matching conditions for $b_n$ and $c_n$ are 
\small
\begin{subequations}
\label{eq.bncn}
\begin{eqnarray}
b_n &=& [k^{2n+4}]\left(- i \, \frac{d-1}{d-2} \, s \, T \left(\omega_\parallel^+ + \omega_{\parallel}^{-}\right) + 2 \, i \, s\, T \omega_\perp \right),\quad \label{eq.bn} \\
c_n &=& [k^{2n+4}]\left(- \frac{k^2}{d-2}  - \frac{d-1}{d-2}\,  \omega_\parallel^+ \, \omega_\parallel^- \right).\label{eq.cn}
\end{eqnarray}
\end{subequations}
\normalsize
The coefficients $a_{n}$, $b_{n}$ and $c_{n}$ are directly related to the  transport coefficients defined in the standard way. Up to third order in gradients one has
\begin{align}
&a_0 = \eta,\,\,\,a_1 = \frac{\eta^2 \, \tau_\pi}{s \, T} - \frac{1}{2}\theta_1,  \quad c_0 = \frac{2\,\eta\, \tau_\pi}{(d-1)\, s\, T},\nonumber \\
&b_0 = \theta_2 - \frac{2 \,(d-3)\,\eta^2\,\tau_\pi}{(d-1)\,s \,T}.  
\end{align}
The explicit relation between hydrodynamic dispersion relations~\eqref{eq.omegahs} and hydrodynamic constitutive relations as encapsulated by \eqref{eq.an} and~\eqref{eq.bncn} will directly lead to our main result on the convergence of the gradient expansion in linearized relativistic hydrodynamics. Its importance stems from the fact that it connects well-studied hydrodynamic dispersion relations as series in small~$k$ with position-space hydrodynamic constitutive relations.

\section{Large order behaviour} The analytic properties of the dispersion relations can be used to constrain the growth of transport coefficients.  We expect that in a microscopic theory which respects relativistic causality, the hydrodynamic dispersion relations $\omega_{\perp}(k)$ and $\omega_{\parallel}^\pm(k)$ have at least one branch-point singularity in the complex $k$-plane. 
This is realized in theories of causal hydrodynamics and holography, and in Appendix \ref{appendix_a} we provide an additional argument in favour of it.
It implies that $\omega_{\perp}(k)$ and~$\omega_{\parallel}(k)$ cannot be polynomials in $k$, 
so the hydrodynamic gradient expansion~\eqref{const_rel_grad_exp} following from the matching conditions~\eqref{eq.an} and~\eqref{eq.bncn} must contain an infinite number of terms. Moreover, 
the transport coefficients $a_n, b_n$ and $c_n$ 
grow geometrically in a manner controlled by the position of the branch points closest to $k = 0$ \cite{wilf2005generatingfunctionology}
\begin{equation}
\lim_{n\to\infty} |a_n|^\frac{1}{n} =  |k^{(A)}_*|^{-2}, \label{generic_formula_growth} 
\end{equation}
where $|k^{(A)}_*|$ denotes the modulus of the branch point location, and analogous expressions hold for $b_{n}$ and $c_{n}$. Note that $|k^{(A)}_*|,|k^{(B)}_*|,|k^{(C)}_*|$ correspond to the closest branch point between $\omega_\perp$ and $\omega_\parallel$ as dictated by \eqref{eq.an} and~\eqref{eq.bncn}. The power 
appearing on the right hand side of \eqref{generic_formula_growth} is due to the fact that the transport coefficients are coefficients of a Taylor series in $k^2$. 

\section{Convergence} The convergence properties of the series~\eqref{const_rel_grad_exp} depend on the behavior of the transport coefficients~$a_{n}$, $b_{n}$ and $c_{n}$ as well as on the particular solution $\epsilon$ and $\textbf{u}$. Here we show that the support in momentum space of the latter plays a crucial role in determining the radius of convergence of the gradient expansion. 
We focus on square-integrable functions, thus excluding trivial cases for which the gradient expansion truncates.

We assume that the flow is homogeneous in the $x^1,...,x^{d-2}$ directions, and define $x \equiv x^{d-1}$. Furthermore, we take the Fourier transforms of $\epsilon(t,x)$ and $u^i(t,x)$, $\hat{\epsilon}(t,k)$ and $\hat{u}^i(t,k)$, to vanish for $|k| > k_{max}$. 
In the linearized regime, the support is time-independent and this condition is a restriction on the initial data. 

According to the Paley-Wiener theorem~\cite{strichartz2003guide}, the Fourier transform of a square-integrable function $\hat{f}(k)$ supported in $|k| \leq k_{max}$ is an entire function of exponential type $k_{max}$.\footnote{See Appendix \ref{appendix_a} for the relevant mathematical background.} For a function of this kind, it follows that \cite{levin1996lectures}
\begin{equation}
\limsup_{n\to\infty} |f^{(n)}(x)|^\frac{1}{n} = k_{max}.  
\end{equation}
Consider now the $A$-contribution to~\eqref{const_rel_grad_exp}. For a compactly-supported $\hat{u}^i$, $\sigma_{jl}(t,x)$ will be of exponential type $k_{max}$ for all times. Hence
\begin{equation}
\limsup_{n\to\infty} |\partial_x^{2n} \sigma_{jl}(t,x)|^\frac{1}{n} = k_{max}^2. 
\end{equation}
Applying the root test results in the following convergence criterion for the $A$-contribution to~\eqref{const_rel_grad_exp}
\begin{equation}
\limsup_{n\to\infty} |a_n \partial_x^{2n} \sigma_{jl}(t,x)|^\frac{1}{n} = \frac{k_{max}^2}{|k^{(A)}_*|^2} < 1. 
\end{equation}
Analogous arguments apply to the remaining pieces of~\eqref{const_rel_grad_exp}. 
Let us define
\begin{equation}
k_* =\operatorname{min}\{|k^{(A)}_*|, |k^{(B)}_*|,|k^{(C)}_*|.
\end{equation}
We are now ready to state our main result for a generic excitation:
\begin{quote}
The gradient expansion is 
a convergent series if the support of the hydrodynamic perturbations and their time-derivatives is smaller than the microscopic momentum scale $k_*$. If the support exceeds $k_*$ then the series is divergent; this includes the case of data which is not compactly supported.
\end{quote}
It follows that if the hydrodynamic series is convergent then its data is compactly supported with support that does not exceed $k_*$.\footnote{In the special case where only the shear channel is excited the appropriate notion of $k_* = |k_*^{(A)}|$.\label{footnote.keqkA}}

Even if divergent, the partial sums of the gradient expansion only grow geometrically as long as the support of the hydrodynamic fields in $k$-space does not extend to infinity. 
If it does, this geometric divergence is enhanced to the factorial one known from the studies of expanding geometries~\cite{Heller:2013fn,Heller:2015dha,Basar:2015ava,Aniceto:2015mto,Heller:2016gbp,Heller:2016rtz,Heller:2018qvh}.

For a flow without any symmetry restrictions, we can argue heuristically that the same convergence conditions hold. Let us focus again on the $A$-contribution to~\eqref{const_rel_grad_exp}. Truncating the series to $N$-th order results in 
\begin{equation}
\begin{split}
-\int_{\mathbb R^{d-1}} d^{d-1}\textbf{k}\,\left[\sum_{n=0}^N a_n \left(\textbf{k}^2\right)^n  \right]\hat\sigma_{ij}(t,\textbf{k}) \, e^{i \, \textbf{k} \cdot \textbf{x}}, \label{master} 
\end{split}
\end{equation}
where we have interchanged the order of summation and integration. According to \eqref{generic_formula_growth}, the partial sums appearing in \eqref{master} are convergent as $N  \to \infty$, provided that they are evaluated at $|\textbf{k}| < |k^{(A)}_*|$. Outside this ($d-1$)-dimensional sphere we get a non-convergent series. Hence, it seems natural to assume that the condition for \eqref{master} to converge as $N \to \infty$ is that the hydrodynamic variable $\hat{\textbf{u}}$ does not have support past $|k^{(A)}_*|$. Analogous arguments would hold also for the $B$- and $C$-pieces, supporting the fact that the convergence criterion spelled out before is fully general. 

\begin{figure}[t!]
\begin{center}
\includegraphics[width=\columnwidth]{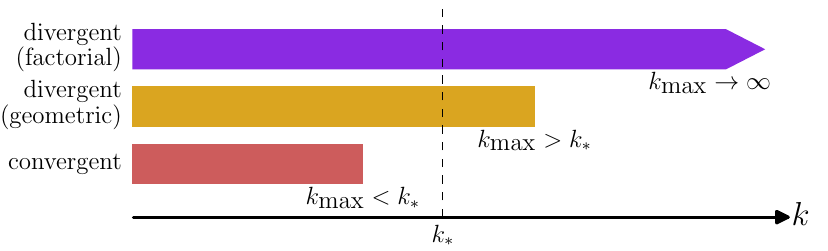}
\caption{The key quantity determining whether the gradient expansion diverges is the support of hydrodynamic fields in momentum space, $k_{\text{max}}$, represented in this figure by the intervals. If the support extends to infinity, the expansion diverges factorially. If it exceeds $k_*$ but truncates at some finite momentum~$k_{max}$, it diverges geometrically. If $k_{max} < k_*$, it converges. See Fig.~\ref{fig:plot_MIS_shear_channel} for an explicit example illustrating this general behavior.
\label{fig.gantt}}
\end{center}
\end{figure}

\section{An illustrative example} For illustration, 
we consider a shear channel perturbation in M{\"u}ller-Israel-Stewart theory~\cite{Muller:1967zza,Israel1976Sep, Israel:1979wp}, 
\begin{equation}
\epsilon = 0,\,\,\,\textbf{u} = (u_1(t,x),0,...,0).
\end{equation} 
We emphasise that our results apply more generally, but this is a particularly simple model of equilibration compatible with relativistic causality featuring a hydrodynamic regime.

In this case, the only tensor structure contributing to \eqref{const_rel_grad_exp} is the shear tensor and the only nontrivial independent component of the constitutive relations is 
\begin{equation}
\Pi_{1,d-1}(t,x) = - \sum_{n=0}^{\infty} a_n (-1)^n \partial_x^{2n+1} u_1(t,x). \label{grad_exp_shear} 
\end{equation}
The $a_n$ transport coefficients can be computed in closed form, since the shear hydrodynamic mode is known exactly~\cite{Baier:2007ix}, 
\begin{equation}
\label{MIS_omega_perp}
\omega_\perp(k) = i \frac{-1+ \sqrt{1- 4 \, D \,\tau_\pi \, k^2}}{2 \, \tau_\pi}, 
\end{equation}
where $D \equiv \eta/(s \, T) = (d-1)/(d-2) \, \Gamma$ is the diffusion constant. M{\"u}ller-Israel-Stewart theory contains also a single nonhydrodynamic shear mode which differs from \eqref{MIS_omega_perp} by the sign of the square root. 
The final result for the $a_n$ coefficients is 
\begin{equation}
a_n = s \, T \, {\cal C}_n D^{n+1} \tau_{\pi}^n, 
\end{equation}
where ${\cal C}_n$ are the Catalan numbers. Therefore,
\begin{equation}
|k_*^{(A)}| = \left(\limsup_{n\to\infty} |a_n|^\frac{1}{n}\right)^{-1/2} = 1/\sqrt{4 \, D \, \tau_{\pi}}, \label{k_*_MIS}
\end{equation}
which is also the location of the branch points of \eqref{MIS_omega_perp}, where the hydrodynamic and the nonhydrodynamic mode collide.  

\begin{figure}[t!]
\begin{center}
\includegraphics[width=\columnwidth]{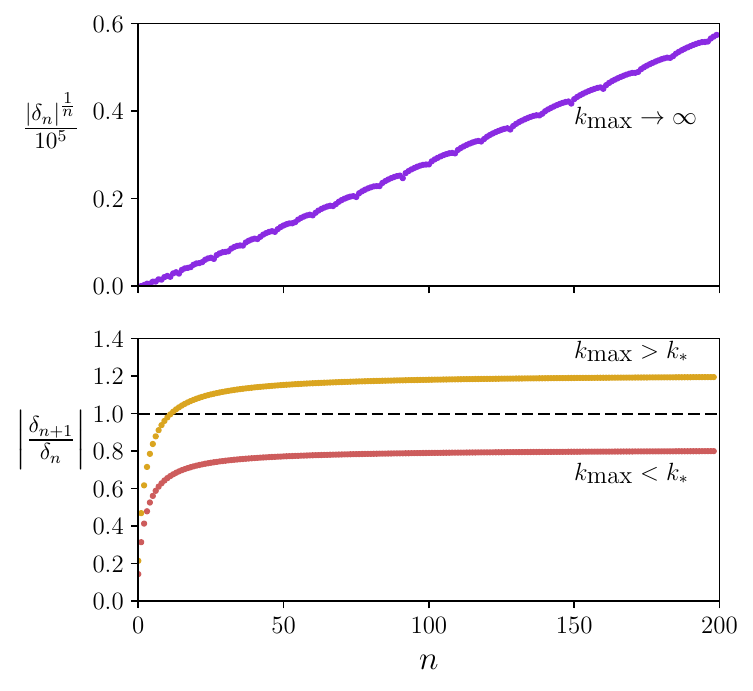}
\caption{\small 
Convergence tests applied to the gradient expansion \eqref{grad_exp_shear}, where $\delta_n$ denotes the  $n$-th contribution. We use $\gamma=0.1$ and consider $t=1$, $x=0.5$ with $s = T = \eta = \tau_\pi = 1$ ($k_*=|k^{(A)}_*|=0.5$, see footnote~\ref{footnote.keqkA}). Datapoint colors and top-bottom ordering correspond to the cases described in Fig.~\ref{fig.gantt}. Upper figure: root test applied to $\delta_n$ when $k_{max} \to \infty$. 
The geometric divergence of the gradient expansion is enhanced to a factorial one. Lower figure: ratio test applied to the solution with $k_{max}=0.55$ (top) and $0.45$ (bottom). The gradient expansion is convergent for $k_{max} < k_*$ and geometrically divergent for $k_{max} > k_*$.  
}
\label{fig:plot_MIS_shear_channel}
\end{center}
\end{figure}

The initial state of the system is fully specified by~$u_1(0,x)$ and~$\partial_{t} u_1(0,x)$. We take $u_1(0,x) = 0$ and
\begin{align}
&\partial_{t}\hat{u}_1(0,k) = \frac{1}{2\pi} e^{-\frac{1}{2}\gamma^2 k^2} \Theta(k_{max}^2- k^2),\label{id_k}
\end{align}
where $\Theta$ is the Heaviside step function. As seen in 
Fig.~\ref{fig:plot_MIS_shear_channel}, the position space gradient expansion is convergent for $k_{max}^2 < 1/(4 \, D \, \tau_{\pi})$, geometrically divergent for $1/(4 \, D \, \tau_{\pi}) \leq k_{max}^2 < \infty$, and factorially divergent for $k_{max} \to \infty$. This is exactly what is  expected on the basis of our general analysis. For more details, see~\cite{heller2020transseries}.

\section{Discussion and outlook} Our work reveals that hydrodynamics itself is neither convergent nor divergent, instead such statements are conditional on the particular solution under consideration.
We have provided a rigorous derivation of a general feature of hydrodynamics which one may phrase heuristically as `hydrodynamics breaks down when gradients become large'. Furthermore we quantified where the hydrodynamic series fails to converge for a general class of models.

Our detailed calculations reveal that the physics governing the convergence of position-space constitutive relations and the convergence of momentum-space dispersion relations are one and the same. In this way we provide a unified perspective on two seemingly disparate lines of research represented by \cite{Heller:2013fn,Buchel:2016cbj,Baggioli:2018bfa,Buchel:2018ttd,Heller:2015dha,Basar:2015ava,Aniceto:2015mto,Denicol:2016bjh,Heller:2016rtz, Heller:2018qvh,Aniceto:2018uik,Denicol:2019lio} and \cite{Withers:2018srf,Grozdanov:2019kge,Grozdanov:2019uhi}.

There are several important lessons that can be drawn from our work in relation to Bjorken flow. Firstly, since we did not impose any particular symmetry, we have shown that the position space hydrodynamic series can still diverge even in the absence of the highly constraining symmetries of boost-invariance. Secondly, while it is conceivable that nonlinear theories diverge due to a factorially growing number of transport coefficients at each order, here we show that this is not a necessary condition since we have at most two transport coefficients at each order and find divergence. Within an analogy with perturbative expansions in quantum mechanical systems dating back to~\cite{Heller:2013fn}, this is similar to renormalon rather than earlier anticipated instanton-related effects~\cite{Beneke:1998ui}. The continuation of these results in the presence of nonlinearities will be discussed in upcoming work \cite{ourupcomingstuff}.

The issue of the convergence of the hydrodynamic gradient expansion is often conflated with the issue of applicability of hydrodynamics for modelling microscopic theories. 
We have shown that the former is determined by support in momentum space, however one can imagine a situation where nonhydrodynamic modes make a significant contribution in a microscopic theory even for states with  support only at low momentum. Such a significant contribution would render hydrodynamics inapplicable even if convergent. 
On the other hand, even if a series diverges it can provide a good description when optimally truncated.
These observations suggest that there is no connection between the two issues at the linear level. It should be noted that this work enables a comprehensive study of optimal truncation -- as a function of initial conditions -- which we leave to future work.

It is very important that complete information about the nonhydrodynamic sector is encoded in the gradient series itself. In the case of an expanding plasma this is very beautifully expressed by the phenomenon of resurgence~\cite{Aniceto:2018bis}, which makes it possible to extract the form of the full solution from the asymptotic  series~\cite{Heller:2015dha,Aniceto:2015mto,Aniceto:2018uik}. An analogous encoding of nonhydrodynamic data in the hydrodynamic sector is seen in the analytic continuation of dispersion relations~\cite{Withers:2018srf}. Generalizations of these ideas based on developments reported in this article are the subject of~\cite{heller2020transseries,ourupcomingstuff}.

\acknowledgements{We would like to thank G.~Dunne, as well as participants of the workshop \emph{Foundational Aspects of Relativistic Hydrodynamics} at Banff International Research Station where this work was presented for the first time for helpful discussions. We also have the pleasure to acknowledge comments on the draft from M.~Baggioli, A.~Buchel, A.~Jansen, M.~Lublinsky, J.~Noronha, C.~Pantelidou. The Gravity, Quantum Fields and Information group at AEI is supported by the Alexander von Humboldt Foundation and the Federal Ministry for Education and Research through the Sofja Kovalevskaja Award. AS and MS are supported by the Polish National Science Centre grant 2018/29/B/ST2/02457. BW is supported by a Royal Society University Research Fellowship.}

\appendix

\section{The dispersion relations and relativistic causality}
\label{appendix_a}

Our main objective in this appendix is to provide additional arguments in favor of the two hypothesis regarding the behavior of the hydrodynamic dispersion relations put forward in the main text: 
\begin{enumerate}
\item $\omega(k)$ has at least one singularity in the complex $k$-plane. 
\item This singularity is a branch point. 
\end{enumerate}

We start by recalling that, under a metric fluctuation $\eta^{\mu\nu} \to \eta^{\mu\nu} + h^{\mu\nu}$, the response of the energy-momentum tensor expectation value in the thermal state is controlled by the retarded two-point function 
\begin{equation}
G^{\mu\nu,\alpha\beta}(t,\textbf{x}) = - i \Theta(t) \langle \left[T^{\mu,\nu}(t,\textbf{x}), T^{\alpha\beta}(0,0) \right] \rangle
\end{equation}
as 
\begin{equation}
\begin{split}
&\delta \langle T^{\mu\nu}(t, \textbf{x}) \rangle = \nonumber \\
&=- \frac{1}{2} \int_{\mathbb R^{1,d-1}} dt'\,d^d\textbf{x}'\,G^{\mu\nu,\alpha\beta}(t-t', \textbf{x}-\textbf{x}') h_{\alpha\beta}(t', \textbf{x}'). \label{response_real_space} 
\end{split}
\end{equation}
The expectation values are taken in the background thermal state. Defining 
\begin{equation}
G^{\mu\nu,\alpha\beta}(t,\textbf{x}) = \int_{\mathbb R^{1,d-1}} d\omega\,d^{d-1}\textbf{k}\, e^{- i \omega t + i \textbf{k}\cdot\textbf{x}} \hat{G}^{\mu\nu,\alpha\beta}(\omega, \textbf{k}), 
\end{equation}
and similarly for $h^{\mu\nu}$, \eqref{response_real_space} can be written as 
\begin{equation}
\begin{split}
&2(2 \pi)^{-d} \delta \langle T^{\mu\nu}(t, \textbf{x}) \rangle = \nonumber \\
&=-\int_{\mathbb R^{1,d-1}} d\omega\,d^{d-1}\textbf{k}\, e^{- i \omega t + i \textbf{k}\cdot\textbf{x}} \hat{G}^{\mu\nu,\alpha\beta}(\omega,\textbf{k}) \hat{h}_{\alpha\beta}(\omega, \textbf{k}). \label{response_momentum_space} 
\end{split}
\end{equation}
Hydrodynamic and nonhydrodynamic frequencies appear as poles of $\hat{G}^{\mu\nu,\alpha\beta}(\omega,\textbf{k})$ which, due to rotational invariance, only depend on $\textbf{k}^2$. See \cite{Kovtun:2005ev} for a detailed discussion of how rotational invariance constrains the form of the retarded correlator. To discuss the interplay between relativistic causality and the analyticity properties of these frequencies, we consider the following setup: we imagine that our metric fluctuation is only active at $t = 0$ and, furthermore, we also assume that it only depends on $x^{d-1} \equiv x$, 
\begin{equation}
h^{\mu\nu}(t,\textbf{x}) = \delta(t) f^{\mu\nu}(x). 
\end{equation}
In momentum space, 
\begin{equation}
\hat{h}^{\mu\nu}(\omega,\textbf{k}) = \frac{1}{2\pi} \delta(k_1)...\delta(k_{d-2}) \hat{f}^{\mu\nu}(k), 
\end{equation}
where we have also defined $k \equiv k_{d-1}$. Hence, 
\begin{equation}
\begin{split}
&2(2\pi)^{1-d}\delta \langle T^{\mu\nu}(t, x) \rangle = \nonumber \\
&=-\int_{\mathbb R^{1,1}} d\omega\,dk \, e^{- i \omega t + i k x} \hat{G}^{\mu\nu,\alpha\beta}(\omega,0,...,0,k) \hat{f}_{\alpha\beta}(k). \label{response_momentum_space_reduced} 
\end{split}
\end{equation}
Performing the integral with respect to $\omega$, we obtain 
\begin{align}
&\delta \langle\hat{T}^{\mu\nu}(t,k)\rangle = \sum_{q=0}^{N_H} \xi^{\mu\nu}_q(k) e^{-i \omega_q(k) t} + \nonumber \\ 
&+ \sum_{q=0}^{N_{NH}} \Xi^{\mu\nu}_q(k) e^{-i \Omega_q(k) t} + \textrm{b.c.} \label{spectral_decomposition}
\end{align}
In writing the spectral decomposition~\eqref{spectral_decomposition}, we have deformed our original integration contour along the real $\omega$-axis to isolate the contributions coming from the singularities of $\hat{G}^{\mu\nu,\alpha\beta}(\omega,k)$ in the lower half of the complex $\omega$-plane. $N_H$, $N_{NH}$ refer respectively to the number of hydrodynamic $\omega_q$ and nonhydrodynamic $\Omega_q$ modes excited by the metric fluctuation, while the excitation coefficients $\xi^{\mu\nu}_q$ and $\Xi^{\mu\nu}_q$ are determined by the residues of the retarded correlator at its poles and the initial data. Finally, $\textrm{b.c.}$ denotes the continuous contributions coming from the branch cuts that might be present. These contributions are absent in theories of causal relativistic hydrodynamics and holography in the semiclassical limit, but do appear in kinetic theory~\cite{Romatschke:2015gic,Kurkela:2017xis}. 

As a final comment about \eqref{spectral_decomposition}, note that we have also assumed that any remaining contribution coming from an integral around infinity can be neglected. This is justified in the case in which our microscopic theory is a Conformal Field Theory and $t > 0$: for $|\omega|\to\infty$ the retarded correlator should reduce to the vacuum result, which does not grow exponentially fast in the same limit. 

Imagine now that $f^{\mu\nu}(x)$ is a square-integrable function supported only for $|x|\leq R$. Relativistic causality demands that, at $t > 0$, the support of $\delta \langle T^{\mu\nu}(t,x)\rangle $ is at most $R+ t$. Let us assume that $\delta \langle T^{\mu\nu}(t,x)\rangle$ is also square-integrable at all times. Then, the Paley-Wiener theorem~\cite{strichartz2003guide} tells us that the spatial Fourier transform of $\delta \langle T^{\mu\nu}(t,x)\rangle $, $\delta \langle T^{\mu\nu}(t,k)\rangle$, is an entire function of exponential type at most $R+t$, also square-integrable along the real $k$-axis. We remind the reader that an entire function $f(z)$ is a function analytic everywhere in the complex $z$-plane, and that an entire function of exponential type $\sigma$ is an entire function obeying the bound 
\begin{equation}
|f(z)| \leq C e^{\sigma |z|},\,\,\, \forall\,z\in\mathbb C,\,\,\, C \in \mathbb R^+. \label{exp_type_bound} 
\end{equation}
In the light of the Paley-Wiener theorem, and when the spectral decomposition~\eqref{spectral_decomposition} holds, property 1 follows by contradiction: if the frequency $\omega(k)$ were entire, its  Laurent series expansion 
\begin{equation}
\omega(k) = \sum_{n=1}^\infty w_n k^n
\end{equation}
would be convergent $\forall\,k\in\mathbb C$, and the bound \eqref{exp_type_bound}, as applied to  $\delta \langle T^{\mu\nu}(t,k)\rangle$, would be violated. This result is in line with the conclusions of \cite{krotscheck1978causality}. Since $\omega_\perp$ is given by a Taylor series in $k^2$, while $\omega_\parallel^\pm$ are series in $k$, the only possible exception to this behavior would be the case in which $\omega_\perp = 0$, $|\omega_\parallel^\pm| \propto |k|$, which corresponds precisely to ideal hydrodynamics. 

On the other hand, property 2 can be justified as follows: if $\omega(k)$ had a pole, $\delta \langle T^{\mu\nu}(t,k)\rangle$ would develop an essential singularity at the pole location, thus failing to be entire. Furthermore, as argued in \cite{krotscheck1978causality}, for systems with a finite number of modes a pole in some dispersion relation entails that the initial value problem does not have a unique solution. 

A final consequence of property 2 is that nonhydrodynamic modes must exist in a theory that respects relativistic causality. These modes, which in principle could be absent if the singularities in the hydrodynamic dispersion relations were poles, appear naturally when analytically continuing these functions past the branch cuts that are actually present. 

\section{The temporal formulation of the hydrodynamic gradient expansion}
\label{appendix_b}

Our objective in this section is to provide a formulation of the gradient-expanded constitutive relations in terms of time derivatives. The elementary tensor structures this gradient expansion is built upon are $\sigma_{ij}$ and $\pi_{ij}^\epsilon$: since we are working in the Landau frame, there is no need to include $\pi^u_{ij}$, as the $t$-component of the conservation equation $\partial_\mu T^{\mu \nu}=0$ allows one to trade $\partial_i u^i$ by $\partial_t \epsilon$. Hence, we write 
\begin{equation}
\Pi_{ij} = - \tilde{A}(\partial_t) \sigma_{ij} - \tilde{B}(\partial_t) \pi_{ij}^\epsilon,
\label{constitutive_relations_temporal}
\end{equation}
where 
\begin{equation}
\tilde{A}(\partial_t) = \sum_{n=0}^{\infty} \tilde{a}_n \partial_t^n, 
\end{equation}
and similarly for $\tilde{B}(\partial_t)$. Here and in the rest of this section the tildes indicate quantities appearing in this temporal gradient expansion.

A matching computation analogous to the one presented in the main text allows one to express the transport coefficients $\tilde{a}_n$ and $\tilde{b}_n$ in terms of the hydrodynamic dispersion relations $k^2_\perp(\omega), k^2_\parallel(\omega)$ of the microscopic theory,  
\begin{align}
&\tilde{a}_n =  s T i^{n+1} [\omega^n]\left(\frac{\omega}{k_\perp^2(\omega)} \right), \label{a_n_temp} \\
&\tilde{b}_n = i^n [\omega^n]\left(\frac{d-1}{d-2}\frac{\omega^2 - \frac{k_\parallel^2(\omega)}{d-1}}{k_\parallel^4(\omega)} - \frac{2 \omega^2}{k_\parallel^2(\omega) k_\perp^2(\omega)} \right).\label{b_n_temp} 
\end{align}
By \eqref{eq.omegahs}, the low-order transport coefficients in this formulation are 
\begin{align}
\tilde{a}_0 = \eta, \quad \tilde{a}_1 = \frac{s T \theta_1}{2\eta}-\eta \tau_\pi, \quad \tilde{b}_0 = \frac{\theta_1}{(d-1)\eta}.
\end{align}
While the hydrodynamic dispersion relations are typically provided by expressing $\omega$ as a function of $k$, the formulation we are considering here has also appeared before in the literature (see \cite{Amado:2007pv,Amado:2008ji} for a discussion in the holographic context). 

We observe that, unlike the purely spatial formulation, the large-order behaviour of the transport coefficients in the temporal gradient expansion is not governed by $\omega(k)$, but by the functions appearing inside the parenthesis in \eqref{a_n_temp} and \eqref{b_n_temp}. In cases where there are no branch points, these functions have poles corresponding to nonhydrodynamic modes at $k=0$ and hence the large-order behavior of the transport coefficients will be controlled by the nonhydrodynamic mode frequency with the smallest absolute value at $k=0$. This indicates that the natural condition for convergence in this case relies on the support of the solution in $\omega$ being below a certain $\tilde{\omega}_*$. One may be tempted to translate this natural formulation from a condition on support in $\omega$ to a condition on support in $k$ for the hydrodynamic fields. However since this step relies on using the dispersion relations the final condition in $k$ will likely be non-universal and potentially not be expressible as a single inequality. This points to the condition in $\omega$ being the most natural formulation for the temporal expansion, and we anticipate other natural conditions arising for other expansions. We leave this issue for future work.\\

The above analysis is illustrated in the following example for M{\"u}ller-Israel-Stewart theory. The dispersion relation $k^2_\perp(\omega)$ is given there by 
\begin{equation}
k_\perp^2(\omega) = \frac{\omega (i + \tau_\pi \omega)}{D}.\label{invdisp} 
\end{equation}
and therefore $a_n = (-1)^n \tau_\pi^n \eta$. Moreover, the sound channel dispersion relation $k_\parallel^2(\omega)$ is such that $\tilde{b}_n = 0$ and hence, in the temporal formulation, only $\sigma_{ij}$ appears in~\eqref{constitutive_relations_temporal}. 
For the shear channel \eqref{invdisp} gives rise to a pole in the function appearing in \eqref{a_n_temp} giving $\tilde{\omega}_* = 1/\tau_\pi$. In this simple example this can be straightforwardly converted into a condition for support in $k$, $|k_{max}| < |\tilde{k}_*|$ where
\begin{equation}
|\tilde{k}_*| =\frac{1}{\sqrt{D\,\tau_\pi}}. \label{k_c_MIS}
\end{equation}
Note that this is proportional to, but distinct from, the critical momentum~$k_{*}$ arising in the purely spatial expansion, given by \eqref{k_*_MIS}. We can analogously define $\omega_*$ which is distinct from $\tilde{\omega}_*$. For $|k_{max}| < |\tilde{k}_*|$, the gradient expansion is convergent but showing this is subtle: unless the nonhydrodynamic mode present in the system is initially turned off, the $n$-th term in the gradient expansion is not exponentially suppressed with the order, but rather decays as a power-law. This is due to the fact that the nonhydrodynamic mode evaluates to $-i \, \tilde{\omega}_*$ at $k=0$.

\bibliography{literature}

\providecommand{\href}[2]{#2}\begingroup\raggedright\begin{thebibliography}{10}

\bibitem{Kovtun:2012rj}
P.~Kovtun, ``{Lectures on hydrodynamic fluctuations in relativistic
  theories},'' \href{http://dx.doi.org/10.1088/1751-8113/45/47/473001}{{\em J.
  Phys. A} {\bfseries 45} (2012) 473001},
  \href{http://arxiv.org/abs/1205.5040}{{\ttfamily arXiv:1205.5040 [hep-th]}}.

\bibitem{Hartnoll:2016apf}
S.~A. Hartnoll, A.~Lucas, and S.~Sachdev, {\em Holographic quantum matter}.
\newblock MIT press, 2018.
\newblock \href{http://arxiv.org/abs/1612.07324}{{\ttfamily arXiv:1612.07324
  [hep-th]}}.

\bibitem{Florkowski:2017olj}
W.~Florkowski, M.~P. Heller, and M.~Spali{\'n}ski, ``{New theories of
  relativistic hydrodynamics in the LHC era},''
  \href{http://dx.doi.org/10.1088/1361-6633/aaa091}{{\em Rept. Prog. Phys.}
  {\bfseries 81} no.~4, (2018) 046001},
\href{http://arxiv.org/abs/1707.02282}{{\ttfamily arXiv:1707.02282 [hep-ph]}}.

\bibitem{Romatschke:2017ejr}
P.~Romatschke and U.~Romatschke,
  \href{http://dx.doi.org/10.1017/9781108651998}{{\em {Relativistic Fluid
  Dynamics In and Out of Equilibrium}}}.
\newblock Cambridge Monographs on Mathematical Physics. Cambridge University
  Press, 5, 2019.
\newblock \href{http://arxiv.org/abs/1712.05815}{{\ttfamily arXiv:1712.05815
  [nucl-th]}}.

\bibitem{Heinz:2013th}
U.~Heinz and R.~Snellings, ``{Collective flow and viscosity in relativistic
  heavy-ion collisions},''
  \href{http://dx.doi.org/10.1146/annurev-nucl-102212-170540}{{\em Ann. Rev.
  Nucl. Part. Sci.} {\bfseries 63} (2013) 123--151},
  \href{http://arxiv.org/abs/1301.2826}{{\ttfamily arXiv:1301.2826 [nucl-th]}}.

\bibitem{Busza:2018rrf}
W.~Busza, K.~Rajagopal, and W.~van~der Schee, ``{Heavy Ion Collisions: The Big
  Picture, and the Big Questions},''
  \href{http://dx.doi.org/10.1146/annurev-nucl-101917-020852}{{\em Ann. Rev.
  Nucl. Part. Sci.} {\bfseries 68} (2018) 339--376},
  \href{http://arxiv.org/abs/1802.04801}{{\ttfamily arXiv:1802.04801
  [hep-ph]}}.

\bibitem{Shibata:2017jyf}
M.~Shibata, K.~Kiuchi, and Y.-i. Sekiguchi, ``{General relativistic viscous
  hydrodynamics of differentially rotating neutron stars},''
  \href{http://dx.doi.org/10.1103/PhysRevD.95.083005}{{\em Phys. Rev. D}
  {\bfseries 95} no.~8, (2017) 083005},
  \href{http://arxiv.org/abs/1703.10303}{{\ttfamily arXiv:1703.10303
  [astro-ph.HE]}}.

\bibitem{Alford:2017rxf}
M.~G. Alford, L.~Bovard, M.~Hanauske, L.~Rezzolla, and K.~Schwenzer, ``{Viscous
  Dissipation and Heat Conduction in Binary Neutron-Star Mergers},''
  \href{http://dx.doi.org/10.1103/PhysRevLett.120.041101}{{\em Phys. Rev.
  Lett.} {\bfseries 120} no.~4, (2018) 041101},
  \href{http://arxiv.org/abs/1707.09475}{{\ttfamily arXiv:1707.09475 [gr-qc]}}.

\bibitem{Bemfica:2019cop}
F.~S. Bemfica, M.~M. Disconzi, and J.~Noronha, ``{Causality of the
  Einstein-Israel-Stewart Theory with Bulk Viscosity},''
  \href{http://dx.doi.org/10.1103/PhysRevLett.122.221602}{{\em Phys. Rev.
  Lett.} {\bfseries 122} no.~22, (2019) 221602},
  \href{http://arxiv.org/abs/1901.06701}{{\ttfamily arXiv:1901.06701 [gr-qc]}}.

\bibitem{Caldarelli:2008mv}
M.~M. Caldarelli, O.~J. Dias, R.~Emparan, and D.~Klemm, ``{Black Holes as Lumps
  of Fluid},'' \href{http://dx.doi.org/10.1088/1126-6708/2009/04/024}{{\em
  JHEP} {\bfseries 04} (2009) 024},
  \href{http://arxiv.org/abs/0811.2381}{{\ttfamily arXiv:0811.2381 [hep-th]}}.

\bibitem{Green:2013zba}
S.~R. Green, F.~Carrasco, and L.~Lehner, ``{Holographic Path to the Turbulent
  Side of Gravity},'' \href{http://dx.doi.org/10.1103/PhysRevX.4.011001}{{\em
  Phys. Rev. X} {\bfseries 4} no.~1, (2014) 011001},
  \href{http://arxiv.org/abs/1309.7940}{{\ttfamily arXiv:1309.7940 [hep-th]}}.

\bibitem{Baier:2007ix}
R.~Baier, P.~Romatschke, D.~T. Son, A.~O. Starinets, and M.~A. Stephanov,
  ``{Relativistic viscous hydrodynamics, conformal invariance, and
  holography},'' \href{http://dx.doi.org/10.1088/1126-6708/2008/04/100}{{\em
  JHEP} {\bfseries 04} (2008) 100},
  \href{http://arxiv.org/abs/0712.2451}{{\ttfamily arXiv:0712.2451 [hep-th]}}.

\bibitem{Kovtun:2005ev}
P.~K. Kovtun and A.~O. Starinets, ``{Quasinormal modes and holography},''
  \href{http://dx.doi.org/10.1103/PhysRevD.72.086009}{{\em Phys.Rev.}
  {\bfseries D72} (2005) 086009},
\href{http://arxiv.org/abs/hep-th/0506184}{{\ttfamily arXiv:hep-th/0506184
  [hep-th]}}.

\bibitem{Heller:2013fn}
M.~P. Heller, R.~A. Janik, and P.~Witaszczyk, ``{Hydrodynamic Gradient
  Expansion in Gauge Theory Plasmas},''
  \href{http://dx.doi.org/10.1103/PhysRevLett.110.211602}{{\em Phys.Rev.Lett.}
  {\bfseries 110} no.~21, (2013) 211602},
\href{http://arxiv.org/abs/1302.0697}{{\ttfamily arXiv:1302.0697 [hep-th]}}.

\bibitem{Buchel:2016cbj}
A.~Buchel, M.~P. Heller, and J.~Noronha, ``{Entropy Production, Hydrodynamics,
  and Resurgence in the Primordial Quark-Gluon Plasma from Holography},''
  \href{http://dx.doi.org/10.1103/PhysRevD.94.106011}{{\em Phys. Rev. D}
  {\bfseries 94} no.~10, (2016) 106011},
  \href{http://arxiv.org/abs/1603.05344}{{\ttfamily arXiv:1603.05344
  [hep-th]}}.

\bibitem{Baggioli:2018bfa}
M.~Baggioli and A.~Buchel, ``{Holographic Viscoelastic Hydrodynamics},''
  \href{http://dx.doi.org/10.1007/JHEP03(2019)146}{{\em JHEP} {\bfseries 03}
  (2019) 146}, \href{http://arxiv.org/abs/1805.06756}{{\ttfamily
  arXiv:1805.06756 [hep-th]}}.

\bibitem{Buchel:2018ttd}
A.~Buchel, ``{Non-conformal holographic Gauss-Bonnet hydrodynamics},''
  \href{http://dx.doi.org/10.1007/JHEP03(2018)037}{{\em JHEP} {\bfseries 03}
  (2018) 037}, \href{http://arxiv.org/abs/1801.06165}{{\ttfamily
  arXiv:1801.06165 [hep-th]}}.

\bibitem{Aniceto:2018uik}
I.~Aniceto, B.~Meiring, J.~Jankowski, and M.~Spali{\'n}ski, ``{The large
  proper-time expansion of Yang-Mills plasma as a resurgent transseries},''
  \href{http://dx.doi.org/10.1007/JHEP02(2019)073}{{\em JHEP} {\bfseries 02}
  (2019) 073}, \href{http://arxiv.org/abs/1810.07130}{{\ttfamily
  arXiv:1810.07130 [hep-th]}}.

\bibitem{Heller:2015dha}
M.~P. Heller and M.~Spali{\'n}ski, ``{Hydrodynamics Beyond the Gradient
  Expansion: Resurgence and Resummation},''
  \href{http://dx.doi.org/10.1103/PhysRevLett.115.072501}{{\em Phys. Rev.
  Lett.} {\bfseries 115} no.~7, (2015) 072501},
  \href{http://arxiv.org/abs/1503.07514}{{\ttfamily arXiv:1503.07514
  [hep-th]}}.

\bibitem{Basar:2015ava}
G.~Ba{\c s}ar and G.~V. Dunne, ``{Hydrodynamics, resurgence, and
  transasymptotics},'' \href{http://dx.doi.org/10.1103/PhysRevD.92.125011}{{\em
  Phys. Rev.} {\bfseries D92} no.~12, (2015) 125011},
\href{http://arxiv.org/abs/1509.05046}{{\ttfamily arXiv:1509.05046 [hep-th]}}.

\bibitem{Aniceto:2015mto}
I.~Aniceto and M.~Spali{\'n}ski, ``{Resurgence in Extended Hydrodynamics},''
  \href{http://dx.doi.org/10.1103/PhysRevD.93.085008}{{\em Phys. Rev.}
  {\bfseries D93} no.~8, (2016) 085008},
\href{http://arxiv.org/abs/1511.06358}{{\ttfamily arXiv:1511.06358 [hep-th]}}.

\bibitem{Denicol:2016bjh}
G.~S. Denicol and J.~Noronha, ``{Divergence of the Chapman-Enskog expansion in
  relativistic kinetic theory},''
  \href{http://arxiv.org/abs/1608.07869}{{\ttfamily arXiv:1608.07869
  [nucl-th]}}.

\bibitem{Heller:2016rtz}
M.~P. Heller, A.~Kurkela, M.~Spali{\'n}ski, and V.~Svensson,
  ``{Hydrodynamization in kinetic theory: Transient modes and the gradient
  expansion},'' \href{http://dx.doi.org/10.1103/PhysRevD.97.091503}{{\em Phys.
  Rev.} {\bfseries D97} no.~9, (2018) 091503},
\href{http://arxiv.org/abs/1609.04803}{{\ttfamily arXiv:1609.04803 [nucl-th]}}.

\bibitem{Heller:2018qvh}
M.~P. Heller and V.~Svensson, ``{How does relativistic kinetic theory remember
  about initial conditions?},''
  \href{http://dx.doi.org/10.1103/PhysRevD.98.054016}{{\em Phys. Rev.}
  {\bfseries D98} no.~5, (2018) 054016},
\href{http://arxiv.org/abs/1802.08225}{{\ttfamily arXiv:1802.08225 [nucl-th]}}.

\bibitem{Denicol:2019lio}
G.~S. Denicol and J.~Noronha, ``{Exact hydrodynamic attractor of an
  ultrarelativistic gas of hard spheres},''
  \href{http://dx.doi.org/10.1103/PhysRevLett.124.152301}{{\em Phys. Rev.
  Lett.} {\bfseries 124} no.~15, (2020) 152301},
  \href{http://arxiv.org/abs/1908.09957}{{\ttfamily arXiv:1908.09957
  [nucl-th]}}.

\bibitem{1963AnPhy..24..419K}
L.~P. {Kadanoff} and P.~C. {Martin}, ``{Hydrodynamic equations and correlation
  functions},'' \href{http://dx.doi.org/10.1016/0003-4916(63)90078-2}{{\em
  Annals of Physics} {\bfseries 24} (Oct., 1963) 419--469}.

\bibitem{Withers:2018srf}
B.~Withers, ``{Short-lived modes from hydrodynamic dispersion relations},''
  \href{http://dx.doi.org/10.1007/JHEP06(2018)059}{{\em JHEP} {\bfseries 06}
  (2018) 059}, \href{http://arxiv.org/abs/1803.08058}{{\ttfamily
  arXiv:1803.08058 [hep-th]}}.

\bibitem{Grozdanov:2019kge}
S.~Grozdanov, P.~K. Kovtun, A.~O. Starinets, and P.~Tadi\'c, ``{Convergence of
  the Gradient Expansion in Hydrodynamics},''
  \href{http://dx.doi.org/10.1103/PhysRevLett.122.251601}{{\em Phys. Rev.
  Lett.} {\bfseries 122} no.~25, (2019) 251601},
  \href{http://arxiv.org/abs/1904.01018}{{\ttfamily arXiv:1904.01018
  [hep-th]}}.

\bibitem{Grozdanov:2019uhi}
S.~Grozdanov, P.~K. Kovtun, A.~O. Starinets, and P.~Tadi\'c, ``{The complex
  life of hydrodynamic modes},''
  \href{http://dx.doi.org/10.1007/JHEP11(2019)097}{{\em JHEP} {\bfseries 11}
  (2019) 097}, \href{http://arxiv.org/abs/1904.12862}{{\ttfamily
  arXiv:1904.12862 [hep-th]}}.

\bibitem{Grozdanov:2015kqa}
S.~Grozdanov and N.~Kaplis, ``{Constructing higher-order hydrodynamics: The
  third order},'' \href{http://dx.doi.org/10.1103/PhysRevD.93.066012}{{\em
  Phys. Rev. D} {\bfseries 93} no.~6, (2016) 066012},
  \href{http://arxiv.org/abs/1507.02461}{{\ttfamily arXiv:1507.02461
  [hep-th]}}.

\bibitem{Diles:2019uft}
S.~M. Diles, L.~A. Mamani, A.~S. Miranda, and V.~T. Zanchin, ``{Third-order
  relativistic hydrodynamics: dispersion relations and transport coefficients
  of a dual plasma},'' \href{http://dx.doi.org/10.1007/JHEP05(2020)019}{{\em
  JHEP} {\bfseries 05} (2020) 019},
  \href{http://arxiv.org/abs/1909.05199}{{\ttfamily arXiv:1909.05199
  [hep-th]}}.

\bibitem{Romatschke:2015gic}
P.~Romatschke, ``{Retarded correlators in kinetic theory: branch cuts, poles
  and hydrodynamic onset transitions},''
  \href{http://dx.doi.org/10.1140/epjc/s10052-016-4169-7}{{\em Eur. Phys. J.}
  {\bfseries C76} no.~6, (2016) 352},
\href{http://arxiv.org/abs/1512.02641}{{\ttfamily arXiv:1512.02641 [hep-th]}}.

\bibitem{Grozdanov:2016vgg}
S.~Grozdanov, N.~Kaplis, and A.~O. Starinets, ``{From strong to weak coupling
  in holographic models of thermalization},''
  \href{http://dx.doi.org/10.1007/JHEP07(2016)151}{{\em JHEP} {\bfseries 07}
  (2016) 151},
\href{http://arxiv.org/abs/1605.02173}{{\ttfamily arXiv:1605.02173 [hep-th]}}.

\bibitem{Bu:2014sia}
Y.~Bu and M.~Lublinsky, ``{All order linearized hydrodynamics from
  fluid-gravity correspondence},''
  \href{http://dx.doi.org/10.1103/PhysRevD.90.086003}{{\em Phys.Rev.}
  {\bfseries D90} no.~8, (2014) 086003},
\href{http://arxiv.org/abs/1406.7222}{{\ttfamily arXiv:1406.7222 [hep-th]}}.

\bibitem{Bu:2014ena}
Y.~Bu and M.~Lublinsky, ``{Linearized fluid/gravity correspondence: from shear
  viscosity to all order hydrodynamics},''
  \href{http://dx.doi.org/10.1007/JHEP11(2014)064}{{\em JHEP} {\bfseries 11}
  (2014) 064}, \href{http://arxiv.org/abs/1409.3095}{{\ttfamily arXiv:1409.3095
  [hep-th]}}.

\bibitem{Lublinsky:2009kv}
M.~Lublinsky and E.~Shuryak, ``{Improved Hydrodynamics from the AdS/CFT},''
  \href{http://dx.doi.org/10.1103/PhysRevD.80.065026}{{\em Phys. Rev. D}
  {\bfseries 80} (2009) 065026},
  \href{http://arxiv.org/abs/0905.4069}{{\ttfamily arXiv:0905.4069 [hep-ph]}}.

\bibitem{Bu:2015ika}
Y.~Bu and M.~Lublinsky, ``{Linearly resummed hydrodynamics in a weakly curved
  spacetime},'' \href{http://dx.doi.org/10.1007/JHEP04(2015)136}{{\em JHEP}
  {\bfseries 04} (2015) 136}, \href{http://arxiv.org/abs/1502.08044}{{\ttfamily
  arXiv:1502.08044 [hep-th]}}.

\bibitem{Bu:2015bwa}
Y.~Bu, M.~Lublinsky, and A.~Sharon, ``{Hydrodynamics dual to
  Einstein-Gauss-Bonnet gravity: all-order gradient resummation},''
  \href{http://dx.doi.org/10.1007/JHEP06(2015)162}{{\em JHEP} {\bfseries 06}
  (2015) 162}, \href{http://arxiv.org/abs/1504.01370}{{\ttfamily
  arXiv:1504.01370 [hep-th]}}.

\bibitem{wilf2005generatingfunctionology}
H.~S. Wilf, ``generatingfunctionology. ak peters,'' 2005.

\bibitem{strichartz2003guide}
R.~S. Strichartz, {\em A guide to distribution theory and Fourier transforms}.
\newblock World Scientific Publishing Company, 2003.

\bibitem{levin1996lectures}
B.~Y. Levin, {\em Lectures on entire functions}, vol.~150.
\newblock American Mathematical Soc., 1996.

\bibitem{Heller:2016gbp}
M.~P. Heller, ``{Holography, Hydrodynamization and Heavy-Ion Collisions},''
  \href{http://dx.doi.org/10.5506/APhysPolB.47.2581}{{\em Acta Phys. Polon.}
  {\bfseries B47} (2016) 2581},
\href{http://arxiv.org/abs/1610.02023}{{\ttfamily arXiv:1610.02023 [hep-th]}}.

\bibitem{Muller:1967zza}
I.~Muller, ``{Zum Paradoxon der Warmeleitungstheorie},''
\href{http://dx.doi.org/10.1007/BF01326412}{{\em Z.Phys.} {\bfseries 198}
  (1967) 329--344}.

\bibitem{Israel1976Sep}
W.~Israel, ``Nonstationary irreversible thermodynamics: a causal relativistic
  theory,'' {\em Annals of Physics} {\bfseries 100} no.~1-2, (1976) 310--331.

\bibitem{Israel:1979wp}
W.~Israel and J.~Stewart, ``{Transient relativistic thermodynamics and kinetic
  theory},''
\href{http://dx.doi.org/10.1016/0003-4916(79)90130-1}{{\em Annals Phys.}
  {\bfseries 118} (1979) 341--372}.

\bibitem{heller2020transseries}
M.~P. Heller, A.~Serantes, M.~Spali{\'n}ski, V.~Svensson, and B.~Withers,
  ``Transseries for causal diffusive systems,''
  \href{http://dx.doi.org/10.1007/JHEP04(2021)192}{{\em Journal of High Energy
  Physics} {\bfseries 2021} no.~4, (Apr., 2021) 192}.

\bibitem{Beneke:1998ui}
M.~Beneke, ``{Renormalons},''
  \href{http://dx.doi.org/10.1016/S0370-1573(98)00130-6}{{\em Phys. Rept.}
  {\bfseries 317} (1999) 1--142},
  \href{http://arxiv.org/abs/hep-ph/9807443}{{\ttfamily arXiv:hep-ph/9807443}}.

\bibitem{ourupcomingstuff}
M.~P. Heller, A.~Serantes, M.~Spali{\'n}ski, V.~Svensson, and B.~Withers, ``{To
  appear},'' 2021.

\bibitem{Aniceto:2018bis}
I.~Aniceto, G.~Basar, and R.~Schiappa, ``{A Primer on Resurgent Transseries and
  Their Asymptotics},''
  \href{http://dx.doi.org/10.1016/j.physrep.2019.02.003}{{\em Phys. Rept.}
  {\bfseries 809} (2019) 1--135},
  \href{http://arxiv.org/abs/1802.10441}{{\ttfamily arXiv:1802.10441
  [hep-th]}}.

\bibitem{Kurkela:2017xis}
A.~Kurkela and U.~A. Wiedemann, ``Analytic structure of nonhydrodynamic modes
  in kinetic theory,''
  \href{http://dx.doi.org/10.1140/epjc/s10052-019-7271-9}{{\em The European
  Physical Journal C} {\bfseries 79} no.~9, (Sept., 2019) 776}.

\bibitem{krotscheck1978causality}
E.~Krotscheck and W.~Kundt, ``Causality criteria,'' {\em Communications in
  Mathematical Physics} {\bfseries 60} no.~2, (1978) 171--180.

\bibitem{Amado:2007pv}
I.~Amado, C.~Hoyos-Badajoz, K.~Landsteiner, and S.~Montero, ``{Absorption
  lengths in the holographic plasma},''
  \href{http://dx.doi.org/10.1088/1126-6708/2007/09/057}{{\em JHEP} {\bfseries
  09} (2007) 057}, \href{http://arxiv.org/abs/0706.2750}{{\ttfamily
  arXiv:0706.2750 [hep-th]}}.

\bibitem{Amado:2008ji}
I.~Amado, C.~Hoyos-Badajoz, K.~Landsteiner, and S.~Montero, ``{Hydrodynamics
  and beyond in the strongly coupled N=4 plasma},''
  \href{http://dx.doi.org/10.1088/1126-6708/2008/07/133}{{\em JHEP} {\bfseries
  07} (2008) 133}, \href{http://arxiv.org/abs/0805.2570}{{\ttfamily
  arXiv:0805.2570 [hep-th]}}.

\end{thebibliography}\endgroup
\bibliographystyle{utphys}

\end{document}